\documentclass[letterpaper]{article}
\usepackage[preprint]{aaai2027}
\usepackage[hyphens]{url}
\usepackage{graphicx}
\graphicspath{{figures/}}
\usepackage{natbib}
\usepackage{caption}
\usepackage{algorithm}
\usepackage{algorithmic}
\usepackage{float}
\usepackage{booktabs}
\usepackage{amsmath,amssymb}

\newcommand{\warranted}{\textsc{Warranted}}
\newcommand{\refuted}{\textsc{Refuted}}
\newcommand{\unknown}{\textsc{Unknown}}
\newtheorem{definition}{Definition}
\newtheorem{proposition}{Proposition}

\title{LASER: Defeasible Certificate Search for Long-Range Protocol Dependencies}
\author{Zihan Tang, Huibiao Zhu\thanks{Corresponding author: hbzhu@sei.ecnu.edu.cn}}
\affiliations{
Shanghai Key Laboratory of Trustworthy Computing, East China Normal University, Shanghai, China\\
Email: 51285902038@stu.ecnu.edu.cn, hbzhu@sei.ecnu.edu.cn
}

\begin{document}
\maketitle

\begin{abstract}
Protocol specifications rarely justify executable behavior in one place: a transition may inherit its source state from a remote control block, name its events only by number, and remain subject to a later correction. Recovering the nearby clause is therefore insufficient; the system must assemble a source-linked certificate and determine whether any stronger rule applies. We introduce LASER, a proof-carrying search method for sparse, defeasible protocol evidence. LASER represents missing support and unresolved corrections as explicit obligations, searches typed evidence routes, and lets an occurrence-aware symbolic checker alone decide whether a certificate warrants, refutes, or leaves a behavior unresolved. On 91 BehaviorCheck families under a matched 2K budget, LASER improves Joint Edge+Proof over the strongest local reader on two model families ($\Delta=+.132$, $p=.017$; $\Delta=+.198$, $p<.001$) while cutting false discoveries by 2--3$\times$. Controlled replays attribute the gain and its limits: neither reference routing nor deterministic citation expansion alone exceeds the local baseline; whole-document reading (29--65K tokens), two-round iterative retrieval, and even oracle minimal-certificate evidence all fail to reach LASER, localizing the bottleneck in typed obligation closure rather than evidence access. An authority-corruption replay shows source-link errors only cause abstention, while the only unsafe-warrant mechanism is mis-recording a Held erratum as Verified, confined to that erratum's family. LASER thus improves long-range dependency recovery while converting incomplete evidence and uncertain authority into explicit residual obligations rather than unsupported behavior.
\end{abstract}

\section{Introduction}

Internet standards are written for exposition and revision, not execution. Consider one real occurrence from RFC~4271 (BGP-4), used throughout this paper:

\begin{quote}\small
\textbf{OPENCONFIRM state.} ``In response to any other event (\emph{events 9, 12--13, 20, 27--28}), the local system: sends a NOTIFICATION message with the Error Code of Finite State Machine Error, \ldots{} and changes its state to Idle.''
\end{quote}

The clause names no event; it names six \emph{numbers}. Events 9, 12, 13, 20, 27, and 28 are defined in a catalog 5{,}518 tokens away. The correct behavior is not one transition but six: \textsc{OpenConfirm} $\xrightarrow{\text{ConnectRetryTimer Expires?}}$ \textsc{Idle}, $\xrightarrow{\text{DelayOpenTimer Expires?}}$ \textsc{Idle}, and so on, each with a mandatory sent NOTIFICATION. A reader that quotes only the local clause can guess ``some event moves OpenConfirm to Idle''; it cannot justify \emph{which} six events, nor bind each to its definition. Worse, a later erratum may replace this text entirely, so even a locally correct clause can be stale.

This paper treats such behavior recovery as \emph{sparse defeasible evidence closure}. Evidence is \emph{sparse} because the required passages occupy a small fraction of the document; \emph{conjunctive} because every typed premise is required; and \emph{defeasible} because a higher-priority correction can overturn local support. A longer context window does not establish closure: failing to retrieve a stronger rule is not evidence that the rule is absent.

\paragraph{Two kinds of obligation.}
Throughout the paper, $O^{+}(h)$ (\emph{support obligations}) are the premises $h$ still needs: a grounded source state, each triggering event, the target state, guards, updates. They are satisfied by \emph{finding} evidence. $O^{-}(h)$ (\emph{defeater obligations}) are the registered rules that could \emph{overturn} $h$: errata, later versions, authority records. They are satisfied by \emph{deciding} them---applicable or not, with authority established. The distinction is directional: absence of evidence leaves $O^+$ open; absence of a decision leaves $O^-$ open, and an open $O^-$ can never certify $h$. This is why LASER abstains \emph{for a stated reason}, instead of merely abstaining with low confidence.

\begin{figure}[H]
  \centering
  \includegraphics[width=\columnwidth]{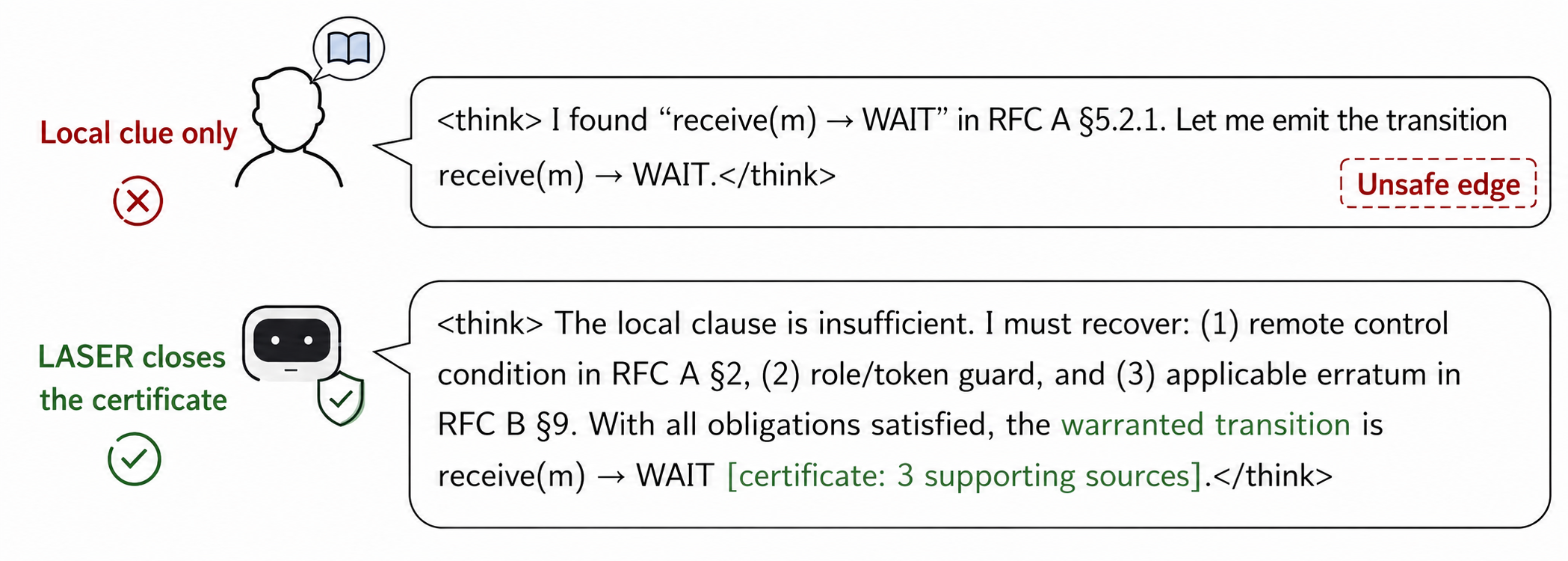}
  \caption{Local evidence versus certificate closure.}
  \label{fig:motivation}
\end{figure}

We first verify the problem in released artifacts: replaying RFCNLP annotations shows transition localization substantially overstates occurrence-aligned edge recovery (Section~\ref{sec:diagnosis}). LASER (\emph{\textbf{L}ong-range \textbf{A}cquisition of \textbf{S}parse \textbf{E}vidence \textbf{R}outes}) then addresses the stricter task with a typed evidence graph and proof-carrying search. A deterministic RuleIR checker alone returns \warranted{}, \refuted{}, or \unknown{}; learned or heuristic policies only order finite, graph-derived actions. Compilation to a timed-automata backend begins only after certificate closure.

Our contributions are:
\begin{itemize}
  \item an artifact-based diagnosis showing that transition localization and occurrence-bound edge recovery diverge;
  \item a source-grounded, three-valued certificate semantics with an explicit treatment of the dual-closure case;
  \item an obligation-aware search algorithm with typed path actions, checker-exclusive verdicts, validated union, and certificate minimality, together with a fairness condition under which relative completeness holds;
  \item a family-paired evaluation with significance tests on two model families, a frozen-completion attribution replay separating routing from citation expansion, long-context, iterative-retrieval, and oracle-evidence controls, a protocol-level cue-transfer audit, and an authority-corruption replay that localizes the only unsafe-warrant mechanism.
\end{itemize}

\section{Related Work}

\paragraph{Protocol knowledge extraction.}
RFCNLP maps annotated protocol text to legacy FSMs and uses those models for attacker synthesis \citep{pacheco2022}. PROSPER uses LLMs to extract protocol artifacts from document chunks \citep{sharma2023}; RFC2Code/APG turns an RFC-derived guidebook into code \citep{long2026}; SecGoal targets formalizable security objectives rather than behavior machines \citep{huang2026}. We reuse RFCNLP artifacts where a source-aligned diagnostic is possible. Their published scores are not placed in one leaderboard because phrase labels, code, security goals, and defeasible behavior certificates use different evaluation objects.

\paragraph{Proof-guided retrieval and verification.}
Closest to our acquisition interface, Aly and Vlachos retrieve documents autoregressively for fact verification and stop when a natural-logic proof over the retrieved set is sufficient \citep{aly2022natural}; ProoFVer executes symbolic natural-logic proofs to return verdicts \citep{krishna2022proofver}. LASER differs in the object being closed: occurrence-bound protocol premises with typed roles, cross-document bindings, and authority-sensitive defeaters whose status can be Verified, Held, or absent. There, evidence sufficiency is judged over a fixed claim and knowledge base; here, the claim itself is a behavior hypothesis whose premises are inherited from remote clauses and can be superseded by later corrections. This difference forces three-valued authority handling and residual-obligation output rather than proof-only verdicts.

\paragraph{Long-context and neuro-symbolic reasoning.}
Long-context models may underuse evidence even when it fits in the input \citep{liu2024,bai2023}. Retrieval-augmented generation supplies external evidence \citep{lewis2020rag}, while multi-hop and adaptive systems interleave retrieval with reasoning \citep{xiong2021,trivedi2023,jeong2024adaptive,sarthi2024raptor}, and agentic methods let a model critique its own retrieval \citep{yao2023react,schick2023toolformer,asai2024selfrag}. These systems improve acquisition, but retrieved relevance alone does not establish that every support and defeater obligation has closed.

Natural-language reasoning benchmarks test rule application, proof production, and first-order inference \citep{clark2020soft,tafjord2021proofwriter,han2024folio}; modular approaches expose intermediate steps or prover actions \citep{creswell2023selection,lyu2023faithful,chen2023pot,yang2023leandojo}, and Logic-LM and LINC delegate entailment to symbolic solvers \citep{pan2023logiclm,olausson2023linc}. LASER preserves the separation of authority but addresses an earlier, open-world problem: the premises must first be acquired across source routes, conjunctions, and possible defeaters. Its acceptance semantics instantiates finite defeasible reasoning \citep{reiter1980,maher2001}. Its \unknown{} decision relates to selective prediction and semantic uncertainty \citep{geifman2017,kuhn2024semantic,wang2023selfconsistency}, but follows explicit residual obligations rather than a confidence threshold.

\section{Diagnosing Sparse Evidence Failures}
\label{sec:diagnosis}

\subsection{From a transition clause to a certificate}

\begin{figure}[t]
  \centering
  \includegraphics[width=\columnwidth,height=0.34\textheight]{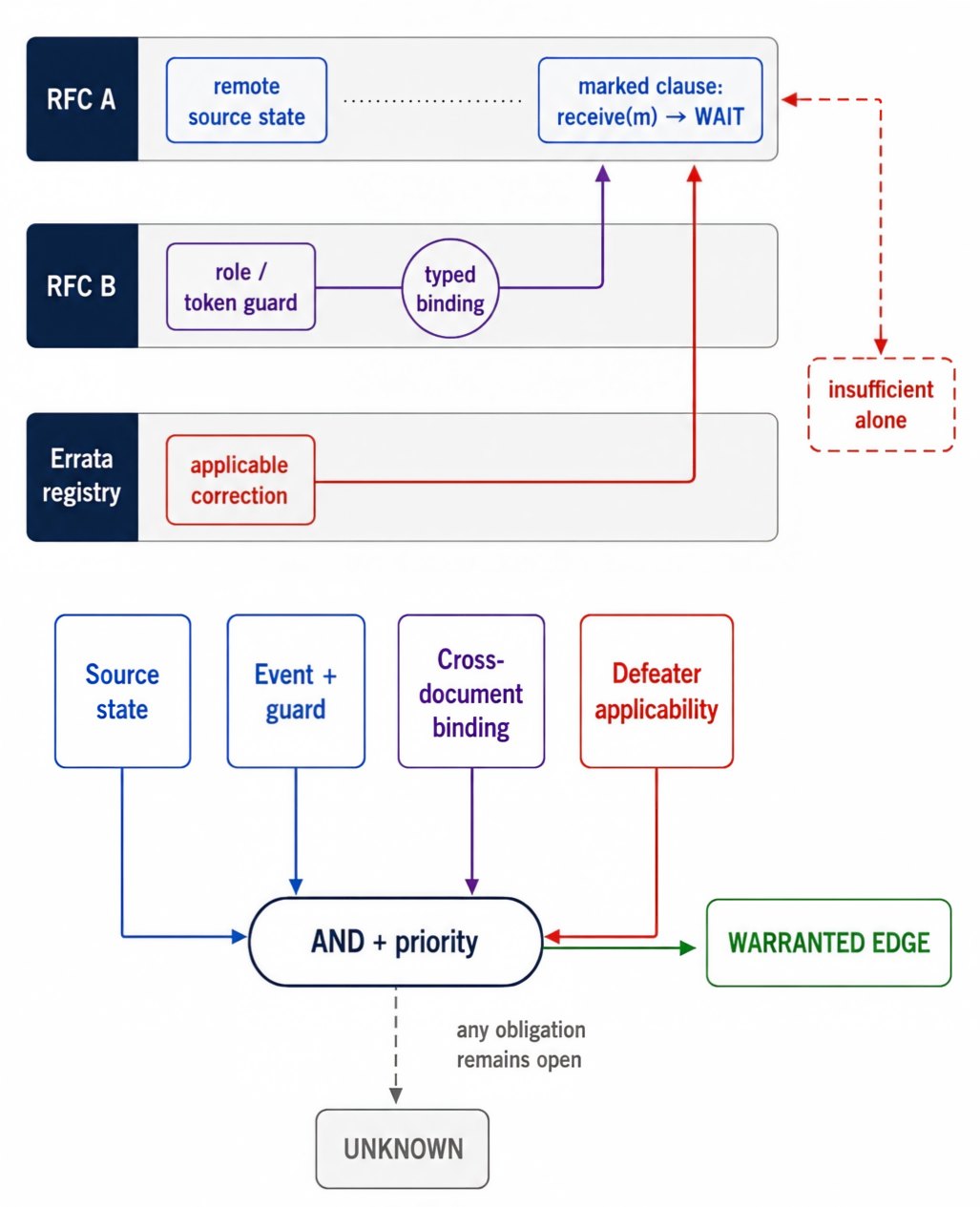}
  \caption{Sparse obligations behind a warranted edge.}
  \label{fig:failure-axes}
\end{figure}

A behavior hypothesis contains a source state, event, target state, participants, guard, update, clock, cross-document bindings, and an applicability audit. Its evidence is not a bag of relevant sentences: it is an AND certificate whose leaves must jointly witness the typed slots and resolve every registered stronger rule. Figure~\ref{fig:failure-axes} makes this closure condition concrete; any open obligation forces \unknown{} rather than an unsupported edge.

\subsection{A released-artifact diagnosis}

We test this distinction on RFCNLP's released annotations and predictions. Predicted phrases are aligned to their source positions and passed through the released transition parser. For TCP, transition localization F1 is .804, but occurrence-aligned edge F1 is .427; for DCCP, .674 and .232. Required evidence occupies only .656\% of RFC tokens at the median, yet the median certificate spans 384 tokens and the maximum 29{,}472. The inherited source state alone is 914 tokens away at the 90th percentile. Locating behavioral language does not recover the sparse control evidence needed to interpret it.

\section{A Complete Certificate, End to End}
\label{sec:walkthrough}

Before defining the semantics, we walk through the OPENCONFIRM example on real data. Figure~\ref{fig:certificate} shows the certificate LASER returns.

\paragraph{Occurrence.} The marked occurrence is the clause ``changes its state to Idle'' inside the OPENCONFIRM block of RFC~4271 (anchor chunk at tokens 26{,}208--26{,}304). Its enclosing control block supplies the source state \textsc{OpenConfirm} and the mandatory sent message NOTIFICATION!.

\paragraph{Open support obligations.} $O^+(h)$ contains one input-event obligation, but the occurrence names it by reference: ``events 9, 12--13, 20, 27--28''. No evidence in the local window can close it. The reference route acquires the event catalog (tokens 19{,}584--20{,}736; up to 5{,}518 tokens away), where Event~9 is ConnectRetryTimer Expires, Event~12 is DelayOpenTimer Expires, Event~13 is IdleHoldTimer Expires, Event~20 is BGPOpen with DelayOpenTimer running, and Events~27--28 are UpdateMsg and UpdateMsgErr.

\paragraph{Defeater obligations.} $O^-(h)$ contains the registered RFC Editor errata for RFC~4271. None binds to this occurrence, so the ledger $\Delta$ resolves each as inapplicable. Had a Verified Technical erratum replaced this clause, the same checker would return \refuted{} for the stale behavior and \warranted{} only for the corrected one; a Held erratum would leave \unknown{}.

\paragraph{Verdict and minimality.} The typed expansion enumerates the six numbers into six atomic input events; the checker grounds each event label in its catalog entry, the source state in the enclosing control, the target in both the anchor chunk and the transition surface, and verifies the mandatory output ordering. All obligations close: the certificate contains 21 source chunks (2{,}016 tokens), and the output is six edges \textsc{OpenConfirm}$\rightarrow$\textsc{Idle} with their grounded inputs. A local reader on the same budget recovers none of the six.

\begin{figure}[t]
  \centering
  \fbox{\parbox{0.97\columnwidth}{\scriptsize
  \textbf{Certificate (abridged).} occurrence: RFC4271 ``changes its state to Idle'' (t26208--26304)\\
  $O^+$: input-event[0] $\leftarrow$ reference ``events 9, 12--13, 20, 27--28''\\
  \quad Event 9 = ConnectRetryTimer Expires (t19584--19680)\\
  \quad Event 12 = DelayOpenTimer Expires; Event 13 = IdleHoldTimer Expires (t19680--19776)\\
  \quad Event 20 = BGPOpen w/ DelayOpenTimer (t20256--20352); Events 27,28 = UpdateMsg, UpdateMsgErr (t20640--20736)\\
  $O^+$: source $\leftarrow$ OPENCONFIRM control block; target $\leftarrow$ occurrence + surface; output $\leftarrow$ NOTIFICATION!\\
  $O^-$: RFC4271 errata $\rightarrow$ no binding to this occurrence\\
  \textbf{Verdict:} \warranted{}; 6 edges; 21 chunks, 2{,}016 tokens; leaf-minimal}}
  \caption{The real certificate for the OPENCONFIRM example: marked occurrence, exact spans, resolved catalog bindings, applicability ledger, and verdict.}
  \label{fig:certificate}
\end{figure}

\section{LASER: Defeasible Certificate Search}

\begin{figure*}[!t]
  \centering
  \includegraphics[width=0.98\textwidth]{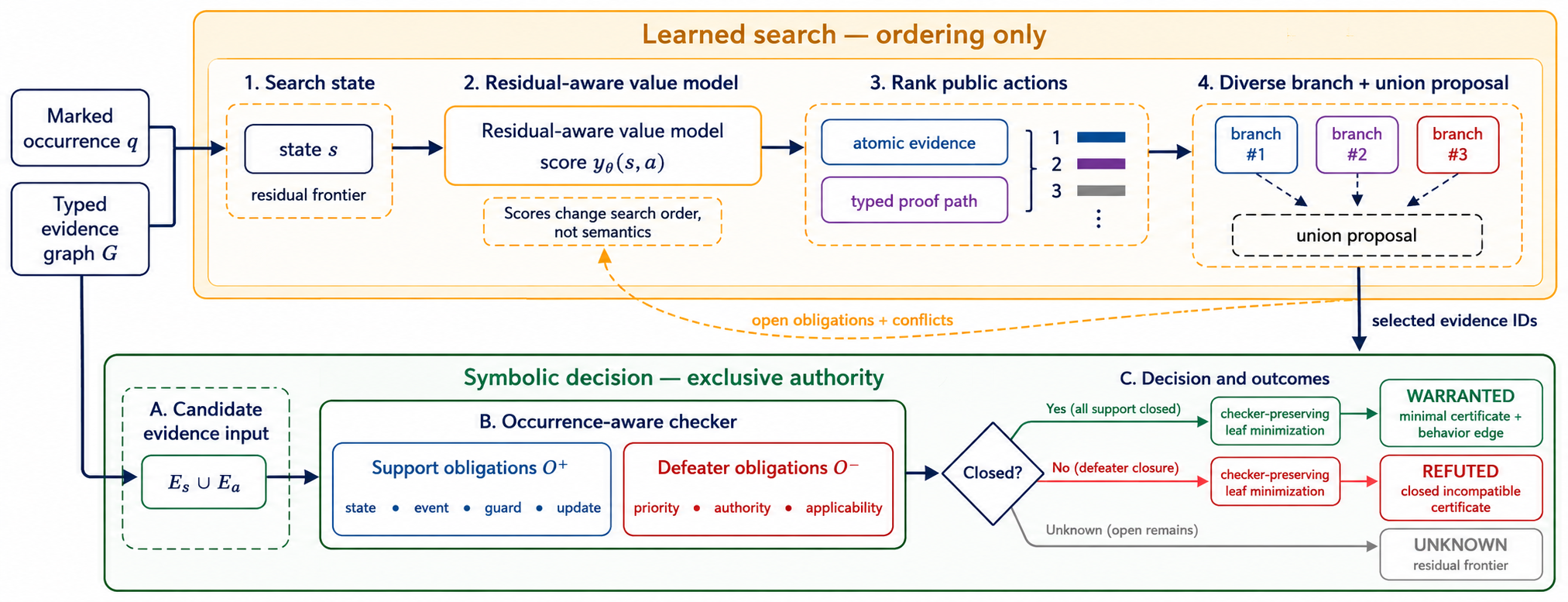}
  \caption{Learned search ordering with checker-exclusive warrant.}
  \label{fig:overview}
\end{figure*}

\paragraph{Input and output.}
LASER receives a marked behavioral occurrence $q$, a public RFC bundle $D$, and a behavior hypothesis $h$. It returns a source-linked certificate for $h$, a certificate for an explicitly incompatible fragment, or \unknown{} with the remaining obligations. The computation has three stages: normalization maps exact RFC spans, citations, section links, and errata into a typed evidence graph; a policy orders finite, graph-derived actions; and after every action a symbolic checker recomputes support, applicability, and conflicts. The verdict depends only on selected source atoms and checker rules; learning affects search order only.

\subsection{Source-grounded semantics}

Let $D$ contain immutable source atoms and their public relations. Each atom $e\in D$ records an exact document/section span and a typed interpretation $(\ell,\tau,v,p)$: literal content, semantic role, version, and priority. The normalizer constructs $G=(V,E_G)$, whose vertices are source atoms and document/section identifiers and whose typed edges encode adjacency, containment, citation, bibliography binding, version, and erratum relations. A behavior hypothesis $h$ has slots for source state, event, target state, participants, guard, update, clock, cross-document binding, and applicability.

\begin{definition}[Obligations]
$O^+(h)$ is the finite set of typed support obligations of $h$. $O^-(h)$ is the set of higher-priority rules whose applicability must be established or ruled out. An evidence set $E$ closes a support obligation only with a source-linked, type-compatible witness \emph{drawn from $E$}; absence from $E$ never closes a defeater.
\end{definition}

Let $\succ$ be an acyclic priority relation supplied by the source policy (e.g., an authoritative correction over its original text). A ledger $\Delta$ maps each $d\in O^-(h)$ to an applicable witness, an inapplicability witness, or \emph{open}. Source-route and erratum bindings must be unique in the full source bundle, not merely within selected evidence. Thus a retrieved correction cannot defeat a clause unless the checker can bind both texts to the same occurrence and establish the correction's authority.

\begin{definition}[Defeasible certificate, evidence-relative]
Given evidence $E$, a certificate $P=(E_P,\Delta)$ for $h$ is \emph{$E$-closed} iff $E_P\subseteq E$, every $o\in O^+(h)$ has a consistent witness in $E_P$, every registered stronger rule is resolved by $\Delta$, and all role, type, clock, version, and binding constraints pass. Write $\mathsf{Closed}(h,E)$ iff some $E$-closed certificate for $h$ exists.
\end{definition}

The restriction $E_P\subseteq E$ is essential: the existential quantifier ranges only over selected evidence, so no conclusion can lean on unseen atoms. Writing $\mathsf{Closed}(x,E)$ as $\mathsf{Cl}(x)$ for readability, the checker implements
\[
\mathsf{dec}(h,E)=
\begin{cases}
\warranted & \mathsf{Cl}(h)\wedge\neg\mathsf{Cl}(\bar h),\\
\refuted & \mathsf{Cl}(\bar h)\wedge\neg\mathsf{Cl}(h),\\
\unknown_{\rm conf} & \mathsf{Cl}(h)\wedge\mathsf{Cl}(\bar h),\\
\unknown & \text{otherwise},
\end{cases}
\]
where $\bar h$ is an explicitly represented incompatible outcome, not an LLM-generated negation. Two clarifications follow. First, witnesses for $h$ and $\bar h$ are drawn from the same $E$, so dual closure is a property of the evidence, not of search order. Second, $\unknown_{\mathrm{conflict}}$ is surfaced as \unknown{} with an explicit conflict marker and both partial certificates; it is not silently resolved. Under acyclic $\succ$ and consistent authority records, a pair cannot both be closed by the same priority-resolved ledger; when authority records genuinely disagree, the conflict case is exactly the intended output.

The checker also returns feedback $z=(v,O^+_{\mathrm{open}},O^-_{\mathrm{open}},\Delta)$, where $v$ is the four-valued internal decision collapsed to three output values. This feedback is the search interface: actions target named residual obligations, while the checker alone removes them. Progress is measured by closed obligations rather than sentence relevance. The guarantee is relative to the finite RuleIR; missing evidence can never be converted into support.

\subsection{Proof-carrying evidence actions}

Figure~\ref{fig:overview} shows the separation. The upper region orders finite public actions and forms union proposals from diverse branches. The lower region recomputes support and defeater closure, minimizes accepted certificates, and returns open obligations otherwise. An action $a=(E_a,k_a,u_a,c_a)$ contains exact evidence IDs $E_a$, an action kind $k_a$, a targeted obligation $u_a$, and token cost $c_a$. It contains no verdict. \emph{Atomic actions} expose one source atom. \emph{Path actions} expose one typed join, such as a source, bibliography binding, and target section. Paths can present mutually dependent leaves in one expansion, but the checker still reconstructs every relation and rejects ambiguous or type-incompatible bindings. For selected evidence $E_s$, applying $a$ means
\[
s\oplus a=\bigl(E_s\cup E_a,\,\mathsf{Check}(h,E_s\cup E_a)\bigr).
\]
Consequently, an incorrect high-ranked action only consumes search budget; it cannot change the acceptance rule.

\subsection{Obligation-aware branch-and-check}

The action abstraction turns certificate completion into branch-and-check. A state $s=(E_s,z_s)$ stores selected evidence and the checker's current feedback. At each expansion, LASER ranks only actions whose evidence is not already selected, creates individually rechecked children, and then creates a merge child from the evidence proposed for different open obligations. One child per residual-obligation type is retained before the remaining beam is filled. This diversity rule prevents several high-scoring support actions from crowding out the only action that investigates an applicable correction. States with the same evidence set are deduplicated; among unresolved states, fewer open obligations, lower evidence cost, and smaller depth define the deterministic queue order.

\begin{algorithm}[t]
\caption{LASER branch-and-check}
\label{alg:laser}
\begin{algorithmic}[1]
\STATE $A\leftarrow\operatorname{TypedActions}(G,h)$
\STATE $z_0\leftarrow\operatorname{Check}(h,\emptyset)$; $Q\leftarrow\{(\emptyset,z_0)\}$
\WHILE{$Q\neq\emptyset$ and budget remains}
  \STATE $s=(E_s,z_s)\leftarrow\operatorname{PopBest}(Q)$
  \IF{$z_s.\mathrm{verdict}=\warranted$}
    \STATE \textbf{return} $\operatorname{Minimize}_{\mathsf{Check}}(h,E_s)$
  \ENDIF
  \STATE $R\leftarrow\operatorname{Rank}_{\theta}(\operatorname{Available}(A,E_s)\mid h,z_s)$
  \STATE $B\leftarrow\operatorname{DiverseTop}(R,z_s.O^+_{\rm open},z_s.O^-_{\rm open})$
  \FOR{$a\in B$}
    \STATE add $s\oplus a$ to $Q$
  \ENDFOR
  \STATE add $\operatorname{CheckUnion}(s,B)$; deduplicate and prune
\ENDWHILE
\STATE \textbf{return} \unknown{} with the best residual frontier
\end{algorithmic}
\end{algorithm}

The evidence budget caps the unique selected-source tokens, the expansion budget caps checker calls to new search nodes, and the beam limits how many unresolved states survive each round. These are search resources, not confidence thresholds. Algorithm~\ref{alg:laser} is written for $h$; $\bar h$ is checked under the same procedure. If neither side closes within budget, LASER returns the lowest-cost frontier with the fewest open obligations, which makes the reason for abstention inspectable.

\subsection{Search values and conservative output}

LASER supports three interchangeable action-ordering policies. \emph{LASER-Symbolic} uses deterministic typed priority, \emph{LASER-Distilled} fits a ridge value to checker-observed progress, and \emph{LASER-LM} asks a fixed language model to rank opaque public action IDs. Each policy receives the same action projections and residual obligations and returns only an ordering. The language model cannot invent a source atom, graph edge, or verdict.

The distilled policy uses process supervision rather than answer supervision. For a development state-action pair, its scalar target is
\[
\begin{aligned}
y_{\lambda}(s,a)={}&\lambda_{\rm close}\,
          \mathbf{1}\!\left[z_{s\oplus a}.v=\warranted\right]\\
        &+\Delta_O(s,a)
        -.02\log(1+c_a)-.01|E_a|,
\end{aligned}
\]
where $\Delta_O(s,a)=|O_{\rm open}(s)|-|O_{\rm open}(s\oplus a)|$. The first term gives priority to a checker-closed child, the second rewards partial closure, and the last two break ties toward shorter evidence. The implementation fixes $\lambda_{\rm close}=8$. This dominance margin is scoped to the observed development statistics: across the 4{,}632 development actions used to fit the value head, the largest single-step obligation reduction is 3 and the largest cost penalty is below .197, so within those bounds 8 makes every checker-closed child outrank any merely partial child. Outside those bounds it remains a fixed lexicographic tie-break, and changing it cannot alter the checker's verdict; it can only alter finite-budget action order.

The ridge features are lexical coverage, action kind, targeted residual type, source distance, and evidence cost. Targets use no expected answer or evaluation certificate. Before returning a proof, a fixed-point deletion pass tests every selected leaf and removes it whenever the same checker still returns \warranted{}. Thus the returned proof is inclusion-minimal, while learning changes only search efficiency and finite-budget recall.

\subsection{Properties}

\begin{proposition}[Conservative soundness]
Relative to sound source links, types, and priority in a finite RuleIR, every LASER \warranted{} output has an $E$-closed certificate; an unresolved stronger rule cannot be treated as absent.
\end{proposition}
\emph{Sketch.} The checker is the sole constructor of a warrant. It quantifies over all registered in-scope higher-priority rules, requires a unique source binding, and restricts every witness to the selected evidence $E$; any failed predicate leaves a residual obligation.

\begin{definition}[Retention fairness]
A search run is \emph{retention-fair} iff, for every state $s$ and action $a$ available at $s$, either some child with evidence $E_s\cup E_a$ enters the queue at some round, or a state with evidence $E_s\cup E_a$ is discarded only in favor of a state with equal or fewer open obligations and no greater cost.
\end{definition}

\begin{proposition}[Relative completeness under fairness]
With complete finite action enumeration, acyclic priority, unbounded budget, and retention fairness, LASER returns a closed certificate iff one exists in the candidate universe.
\end{proposition}
\emph{Sketch.} Under retention fairness, no assignment containing a closing certificate is permanently discarded, so exhaustive expansion visits it; the deletion fixed point preserves warrant.

\emph{Scope of Proposition~2.} The implemented beam and DiverseTop rules are \emph{not} retention-fair in general: a finite beam can discard a state that a later closure would need. We therefore claim completeness only for the exhaustive, fair variant, and treat the implemented algorithm as an anytime procedure whose finite-budget misses are precisely the \unknown{} residuals it reports. This is consistent with the conservative-output contract: a budget-limited \unknown{} never asserts the absence of a certificate.

\begin{proposition}[Minimality]
Every returned certificate is leaf-minimal under the checker: removing any remaining leaf destroys warrant.
\end{proposition}
\emph{Sketch.} The deletion pass reaches a fixed point; by construction every surviving leaf was tested and found necessary. This is inclusion minimality, not globally minimum token cost.

\section{Experiments}
\label{sec:experiments}

\begin{table*}[t]
\centering
\small
\setlength{\tabcolsep}{2.5pt}
\renewcommand{\arraystretch}{1.00}
\begin{tabular*}{\textwidth}{@{\extracolsep{\fill}}lrrrrrrrrrr@{}}
\toprule
Reader & \multicolumn{5}{c}{Edge recovery} & \multicolumn{3}{c}{Proof quality} & \multicolumn{2}{c}{Decision}\\
\cmidrule(lr){2-6} \cmidrule(lr){7-9} \cmidrule(lr){10-11}
 & P & R & $F_1$ & Macro $F_1$ & Exact & Evid. $F_1$ & nDCG & Joint & FDR $\downarrow$ & Abstain\\
\midrule
\multicolumn{11}{l}{\textit{Local context}}\\[-2pt]
Previous context & .647 & .219 & .327 & .302 & .297 & .406 & .412 & .220 & .353 & .538\\
BM25 + source neighbors & .581 & .285 & .382 & .313 & .308 & .486 & .474 & .242 & .419 & .462\\
Occurrence-centered context & .582 & .258 & .358 & .374 & .363 & .559 & .554 & .308 & .418 & .407\\
\addlinespace[2pt]
\multicolumn{11}{l}{\textit{Long-range evidence routing}}\\[-2pt]
LASER (reference) & \textbf{.821} & .609 & .700 & .618 & .593 & .684 & .623 & .407 & \textbf{.179} & .242\\
LASER (hybrid) & .813 & .576 & .674 & \textbf{.630} & \textbf{.626} & \textbf{.692} & \textbf{.646} & \textbf{.440} & .187 & \textbf{.242}\\
LASER (cue-gated) & .810 & \textbf{.623} & \textbf{.704} & .604 & .604 & .668 & .584 & .418 & .190 & .286\\
\bottomrule
\end{tabular*}
\caption{Behavior recovery under a matched 2K evidence budget.}
\label{tab:main-results}
\end{table*}

\begin{figure*}[!t]
  \centering
  \includegraphics[width=0.98\textwidth]{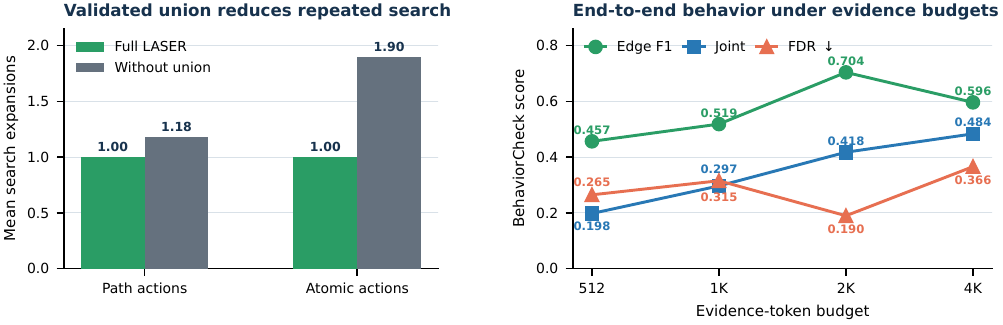}
  \caption{Component ablation and evidence-budget response.}
  \label{fig:conservative-audit}
\end{figure*}

\begin{table*}[t]
\centering
\small
\setlength{\tabcolsep}{5.0pt}
\renewcommand{\arraystretch}{1.04}
\begin{tabular*}{\textwidth}{@{\extracolsep{\fill}}llrrl@{}}
\toprule
Removed component & Evaluation & $\Delta$ correct/closed & $\Delta$ expansions & Safety effect\\
\midrule
Proof paths & 25 applicable errata & $-25$ Closed@1 & $+0.87$ & No unsafe change\\
Checker feedback & 277 controlled routes & $0$ & $0.00$ & 0/167 false routes\\
Diverse beam & 277 controlled routes & $0$ & $0.00$ & 0/167 false routes\\
Validated union & 277 controlled routes & $0$ & $+0.18$ path; $+0.90$ atomic & 0/167 false routes\\
Authority checking & 20 paired families & $-16$ correct claims & N/A & $+8$ unsafe warrants\\
\bottomrule
\end{tabular*}
\caption{Removal ablation across controlled evaluations.}
\label{tab:cross-suite}
\end{table*}

\begin{table*}[t]
\centering
\small
\setlength{\tabcolsep}{4.2pt}
\renewcommand{\arraystretch}{1.02}
\begin{tabular*}{\textwidth}{@{\extracolsep{\fill}}lrrrrrrrr@{}}
\toprule
Slice & $n$ & \multicolumn{3}{c}{Edge $F_1$} & \multicolumn{2}{c}{Exact match} & \multicolumn{2}{c}{Joint Edge+Proof}\\
\cmidrule(lr){3-5} \cmidrule(lr){6-7} \cmidrule(lr){8-9}
 & & Local & LASER & $\Delta$ & Local & LASER & Local & LASER\\
\midrule
\multicolumn{9}{l}{\textit{Dependency regime}}\\[-2pt]
Remote reference & 28 & .025 & \textbf{.726} & +.701 & .000 & \textbf{.643} & .000 & \textbf{.321}\\
Local route & 63 & .579 & \textbf{.688} & +.109 & .444 & \textbf{.587} & .349 & \textbf{.460}\\
\addlinespace[2pt]
\multicolumn{9}{l}{\textit{Protocol}}\\[-2pt]
BGPv4 & 52 & .188 & \textbf{.780} & +.592 & .173 & \textbf{.692} & .096 & \textbf{.404}\\
DCCP & 8 & .769 & .769 & .000 & .625 & .625 & .625 & .625\\
TCP & 31 & .568 & .568 & .000 & .452 & .452 & .387 & .387\\
\bottomrule
\end{tabular*}
\caption{Behavior recovery by dependency regime and protocol.}
\label{tab:dependency-results}
\end{table*}

\subsection{Datasets and Tasks}

\paragraph{BehaviorCheck.}
We derive 103 occurrence-bound behavior families from released BGPv4, DCCP, SCTP, and TCP annotations. Twelve families support task and prompt development; the remaining 91 form the evaluation split. SCTP contributes 13 source transitions, all of which the occurrence-aware projection excludes (10 lack an inheritable source state, three lack a required slot), so the evaluation split contains BGPv4 (52), TCP (31), and DCCP (8) families; SCTP remains in the projection audit. Each family requires an atomic source, event, and target edge together with the evidence that warrants it. Twenty-eight families require a numbered or complementary reference outside the local transition context.

\paragraph{Defeater and authority evaluation.}
DefeaterCheck contains 31 real RFC Editor erratum routes. Twenty-five have one applicable Verified Technical correction; six remain unresolved. A separate authority evaluation contains 20 paired stale and corrected claims (40 decisions) from six protocol families: 12 Verified and 8 Held. We also use 277 controlled long-range routes (110 positive, 167 negative) to measure search behavior under fixed distractors.

\paragraph{Executable backend.}
Closed CoAP timeout certificates are compiled to a timed-automata interface and checked against implementation traces and UPPAAL queries. This case study evaluates the compilation boundary, not natural-language evidence labels.

\subsection{Experimental Settings}

All behavior readers use the same language model, candidate anchor, structured output contract, occurrence-aware checker, and 2K evidence budget. The strongest local baseline on Joint Edge+Proof is the occurrence-centered reader; the strongest on micro Edge F1 is BM25 + source neighbors; both appear in Table~\ref{tab:main-results}. LASER adds two fixed routing policies: a cue-gated policy uses reference search only when a public dependency cue is present, while a hybrid policy combines local and reference evidence. On errata, atomic and path actions share the same checker. We report micro and macro Edge F1, exact edge-set match, evidence F1 and nDCG, Joint Edge+Proof, false discovery rate (complement of precision, reported for continuity), abstention, three-valued accuracy, and unsafe warrants. Prompts, search limits, and remaining hyperparameters are in the supplementary material; every run records prompt and completion hashes.

\subsection{Main Results}

Table~\ref{tab:main-results} reports the matched behavior comparison. Against the strongest local Joint reader (occurrence-centered, .308), the LASER hybrid reaches .440 Joint Edge+Proof; against the strongest local Edge-F1 reader (BM25 + source neighbors, .382), the cue-gated policy reaches .704. Family-paired statistics appear in Table~\ref{tab:paired-stats}: the hybrid improves Joint by $+.132$ (95\% CI $[+.033,+.231]$, permutation $p=.017$; Wilcoxon $p=.011$), exact match by $+.264$ and Edge F1 by $+.256$ (both $p<10^{-3}$). The cue-gated and reference policies improve Edge F1 and exact match significantly ($p<.005$) with positive but marginal Joint deltas ($p=.096$ and $.121$), reflecting the smaller hybrid edge-set improvements on local-route families.

Table~\ref{tab:dependency-results} localizes the gain. On 28 reference-required families, Edge F1 rises from .025 to .726, exact match from .000 to .643, and Joint from .000 to .321. Throughout, ``Local'' denotes the BM25 + source-neighbors reader and ``LASER'' the cue-gated configuration of Table~\ref{tab:main-results}. BGPv4 improves from .188 to .780 Edge F1; DCCP and TCP are unchanged because the cue gate retains local reading when no remote route is exposed.

\begin{table}[t]
\centering\small
\begin{tabular*}{\columnwidth}{@{\extracolsep{\fill}}lcccc@{}}
\toprule
Routing & Expansion & Edge $F_1$ & Joint & Remote $F_1$\\
\midrule
Local & off & .382 & .242 & .025\\
Local & on & .382 & .242 & .025\\
Reference & off & .396 & .253 & .054\\
Reference & on & .700 & .407 & .719\\
\bottomrule
\end{tabular*}
\caption{Attributing the remote-reference gains by replaying frozen completions through the same checker. Neither deterministic citation expansion without routing nor routing without the typed expansion closes remote references; the gain requires both components. Remote column: Edge F1 on the 28 reference-required families.}
\label{tab:attribution}
\end{table}

\subsection{What Produces the Gain? An Attribution Replay}
\label{sec:attribution}

A reviewer can reasonably ask whether the remote-reference gains come from obligation-aware routing or merely from a deterministic citation-expansion rule. We therefore replay \emph{frozen completions} through the same checker in a $2\times2$ design (Table~\ref{tab:attribution}): reference routing on/off $\times$ deterministic expansion on/off. Applying the expansion rule to the local reader's evidence changes nothing (Edge F1 stays .382, remote .025): the rule cannot fire because the event catalog is not in the budget. Conversely, routing without the typed expansion reaches only .396 overall and .054 remote: the model sees the catalog but does not reliably enumerate the referenced numbers. Only the combination closes remote references (.700 overall, .719 remote). Neither component alone differs significantly from the local baseline (Table~\ref{tab:paired-stats}). The gain is therefore attributable to the interaction---routing acquires the typed catalog evidence, and the checker-side expansion closes the numeric obligation---not to either mechanism in isolation.

\begin{table}[t]
\centering\small
\begin{tabular*}{\columnwidth}{@{\extracolsep{\fill}}lcccc@{}}
\toprule
Method & $\Delta$Joint & CI95 & $p$ & $\Delta$Edge\\
\midrule
Hybrid & +0.132$^{*}$ & [0.033,0.231] & 0.017 & +0.256\\
Cue-gated & +0.110 & [-0.011,0.220] & 0.096 & +0.230\\
Reference & +0.099 & [-0.011,0.209] & 0.121 & +0.244\\
Expansion only & -0.066 & [-0.143,0.011] & 0.182 & -0.061\\
Routing only & -0.055 & [-0.132,0.022] & 0.262 & -0.023\\
\bottomrule
\end{tabular*}
\caption{Family-paired LASER variants and attribution arms against the strongest local reader (occurrence-centered) on Joint Edge+Proof, 20{,}000 bootstrap and permutation rounds, family as the unit ($n{=}91$); $p$: two-sided permutation test, $^{*}p<0.05$.}
\label{tab:paired-stats}
\end{table}

\subsection{Cue-Gate Transfer Across Protocols}

Table~\ref{tab:cue-transfer} audits the public cue gate per protocol. The gate is a fixed, prediction-blind pattern test (numbered event reference or explicit complement); gold labels are used only to evaluate it. It fires on all 52 BGPv4 families (recall 1.0), on one of two required DCCP families, and never on TCP, missing one required TCP complement variant. Overall recall is .93 with precision .49: the gate over-fires on BGPv4, where reference routing never removes a correct local edge, and its misses are exactly two uncovered complement phrasings. This delimits where the current routing policy transfers: pattern-level generalization, with recall bounded by public cue coverage.

\begin{table}[t]
\centering\small
\begin{tabular*}{\columnwidth}{@{\extracolsep{\fill}}lccccc@{}}
\toprule
Protocol & Fam. & Fired & Req. & Prec. & Rec.\\
\midrule
BGPv4 & 52 & 52 & 25 & .48 & 1.00\\
DCCP & 8 & 1 & 2 & 1.00 & .50\\
TCP & 31 & 0 & 1 & .00 & .00\\
\midrule
All & 91 & 53 & 28 & .49 & .93\\
\bottomrule
\end{tabular*}
\caption{Public cue-gate transfer audit; Req.: reference required, Prec./Rec.: cue precision/recall. The gate is a fixed, prediction-blind pattern test; gold is used only to evaluate it. It misses two required references (one DCCP and one TCP complement variant) and over-fires on BGPv4, where reference routing never removes a correct local edge.}
\label{tab:cue-transfer}
\end{table}

\subsection{Second-Model Replication, Long-Context, Iterative, and Oracle Controls}
\label{sec:multimodel}

Table~\ref{tab:multimodel} reports a full second-model replication (GLM-5.3, all 91 families, same prompt, checker, and 2K budget) with four additional controls. First, the routing-over-local ordering replicates with larger margins than on the primary model: LASER hybrid reaches .803 Edge F1 / .473 Joint versus .447 / .275 for the strongest local reader (family-paired $\Delta$Joint $=+.198$, $p<.001$; only 2 of 91 families regress). Second, \emph{whole-document reading is the weakest reader of all} (.276 Edge F1, .545 false discovery rate) despite 29K--65K tokens of evidence: unrestricted context adds distractors without telling the model \emph{which} obligations to close. Third, a two-round checker-residual-driven iterative reader under the same total budget does not help (.264), showing that generic adaptive retrieval is not the missing ingredient. Fourth, an \emph{oracle} arm that selects the gold minimal-certificate evidence (labels never shown) reaches only .496: the minimal certificate does not include the catalog chunks that typed expansion needs, so even perfect retrieval of the recorded certificate cannot match routing that acquires the catalog. Together these controls localize the remaining bottleneck precisely where LASER acts---typed obligation closure---rather than in evidence access, budget, or model scale.

\begin{table*}[t]
\centering\small
\begin{tabular*}{\textwidth}{@{\extracolsep{\fill}}lcccccc@{}}
\toprule
Reader (GLM-5.3, $n{=}91$) & Evidence & Edge $F_1$ & Joint & Exact & $\Delta$Joint & $p$\\
\midrule
Iterative retrieval (2 rounds) & 2K & .264 & .176 & .242 & -0.099 & 0.065\\
Direct-full (whole document) & 29--65K & .276 & .209 & .253 & -0.066 & 0.240\\
Local BM25 + source neighbors & 2K & .406 & .275 & .319 & +0.000 & 1.000\\
Occurrence-centered (strongest local) & 2K & .447 & .275 & .352 & --- & ---\\
Oracle certificate evidence & $\leq$2K & .496 & .330 & .407 & +0.055 & 0.474\\
LASER (reference) & 2K & .799 & .484 & .626 & +0.209 & $<$.001\\
LASER (cue-gated) & 2K & .791 & .473 & .626 & +0.198 & $<$.001\\
LASER (hybrid) & 2K & .803 & .473 & .626 & +0.198 & $<$.001\\
\bottomrule
\end{tabular*}
\caption{Full second-model replication on GLM-5.3 (all 91 families; same prompt, checker, and 2K budget). $\Delta$Joint and $p$ are family-paired against the strongest local reader. Whole-document reading, two-round iterative retrieval, and even oracle minimal-certificate evidence do not reach LASER: the bottleneck is typed obligation closure, not evidence access. Oracle uses private certificate evidence for selection only.}
\label{tab:multimodel}
\end{table*}

\subsection{Sensitivity to Authority and Source-Link Errors}
\label{sec:corruption}

Proposition~1 is relative to correct links, types, and priorities, so we measure what happens when that assumption fails. Table~\ref{tab:corruption} replays the deterministic laser-path method on the 20-family authority suite while corrupting one erratum record at a time. The clean replay reproduces all 40 expected labels. Three findings follow. (1)~Corrupting source links (RFC binding or stale text) \emph{never} creates an unsafe warrant; it only converts closures into \unknown{}, so the failure mode is conservative abstention. (2)~The only corruption that creates unsafe warrants is recording a Held erratum as Verified; all 16 unsafe warrants come from the 8 Held families, each corruption affecting exactly its own family. (3)~Flipping Verified to Held only loses closures (24), again without unsafe warrants. End-to-end soundness therefore decomposes: representation errors degrade to abstention, and authority-status errors are the single, localized unsafe-warrant channel.

\begin{table}[t]
\centering\small
\begin{tabular*}{\columnwidth}{@{\extracolsep{\fill}}lccc@{}}
\toprule
Corruption & Records & Unsafe & Lost\\
\midrule
Held $\rightarrow$ Verified & 8 & 16 & 0\\
Verified $\rightarrow$ Held & 12 & 0 & 24\\
RFC binding corrupted & 20 & 0 & 24\\
Stale text corrupted & 20 & 0 & 24\\
\bottomrule
\end{tabular*}
\caption{Joint authority/source-link corruption replay on the 20-family sealed authority suite (40 claims). Clean replay reproduces all 40 expected labels. Unsafe: unsafe warrants; Lost: lost closures. Source-link and binding corruption never creates an unsafe warrant; it degrades to \unknown{}. Mis-recording a Held erratum as Verified is the only unsafe-warrant mechanism, and it affects exactly the corrupted family.}
\label{tab:corruption}
\end{table}

\subsection{Ablation Study}

Table~\ref{tab:cross-suite} reports matched removal effects. Removing proof paths loses all 25 one-expansion closures; removing validated union increases search cost, especially for atomic actions. Checker feedback and beam diversity show no measurable effect on the controlled routes; this is expected rather than anomalous. On those routes the symbolic priority is already near-ceiling: on the 277-route controlled suite, the symbolic policy alone reaches .991 positive certificate recall with zero model calls and a mean of 55 evidence tokens, leaving learned ranking nothing to improve. The situation differs on BehaviorCheck, where the routing policy itself---which evidence family to consult---carries the gain (Section~\ref{sec:attribution}). The practical benefit of learning is thus scoped: it can only reorder finite actions, so it helps exactly where the symbolic order is suboptimal, and our current suites bound that regime. Authority checking is essential for safety: removing it loses 16 correct claim decisions and turns all eight Held corrections into unsafe warrants.

Figure~\ref{fig:conservative-audit} connects the search ablation to the end-to-end budget response. Moving from 512 tokens to 2K recovers missing support; at 4K, additional proof leaves improve Joint Edge+Proof while distractor edges degrade Edge F1 and precision. LASER benefits from targeted acquisition rather than an unrestricted context window.

\subsection{RuleIR Construction Audit}
\label{sec:ruleir}

The BehaviorCheck projection audit covers 162 source transitions: 103 are included and 59 are excluded with machine-checkable reasons (25 ambiguous or missing target, 18 missing source state, 9 missing input event, 3 no control context, 4 requiring review). On the sealed authority suite, all 20 erratum stale texts have a unique occurrence in their frozen RFC, so the binding-uniqueness precondition holds. In the reference run, the typed resolver finds a complete public catalog for 47 of 49 numbered-reference families (the 2 incomplete cases abstain), plus 1 complement family. These audits quantify how the finite RuleIR is built and where it abstains; they do not replace independent annotation review, which remains a stated limitation.

\paragraph{Checker robustness.}
Table~\ref{tab:checker-sensitivity} replays frozen GLM completions through the occurrence checker while sweeping its window constants and removing its alias table. Results saturate at $\pm$192/96 tokens and are unchanged by the aliases or a doubled window; even a four-fold tighter window keeps precision at .880 while losing recall. The checker's constants are therefore not hidden tuning: below saturation it degrades conservatively, and above saturation they are inert.

\begin{table}[t]
\centering\small
\begin{tabular*}{\columnwidth}{@{\extracolsep{\fill}}lcccc@{}}
\toprule
Window & Warranted & Edges & Edge $F_1$ & Joint\\
\midrule
$\pm$64/32 & 59 & 92 & .667 & .418\\
$\pm$128/64 & 67 & 121 & --- & ---\\
$\pm$192/96 & 68 & 122 & .799 & .484\\
$\pm$384/192 & None & None & .799 & .484\\
$\pm$768/384 & 68 & 122 & --- & ---\\
no aliases & 68 & 122 & .799 & .484\\
\bottomrule
\end{tabular*}
\caption{Occurrence-checker sensitivity replay over frozen GLM completions; the deployed window is $\pm$384/192. Results saturate at $\pm$192/96 and are unchanged by the alias table or a doubled window; halving the window to $\pm$64/32 loses recall while keeping precision (.880), i.e., degradation is conservative.}
\label{tab:checker-sensitivity}
\end{table}

\subsection{Analysis}

\paragraph{The gains follow dependency structure.}
The improvement concentrates on reference-required families and the protocol ordering follows measured control distance, supporting the claim that LASER repairs sparse remote dependencies rather than uniformly changing predictions.

\paragraph{Search and decision authority remain separate.}
Paths and validated union reduce the expansions needed to assemble a conjunction; all compared policies share the same occurrence, type, priority, and authority checks, which explains why finite-budget closure improves without increasing unsafe warrants.

\paragraph{Executable interface.}
The compiled CoAP model passes two pinned traces, two multicast queries, and four terminal-aware UPPAAL queries. Query results come from verifier logs; the case study validates the compilation boundary, not natural-language interpretation.

\begin{figure}[!t]
  \centering
  \includegraphics[width=\columnwidth]{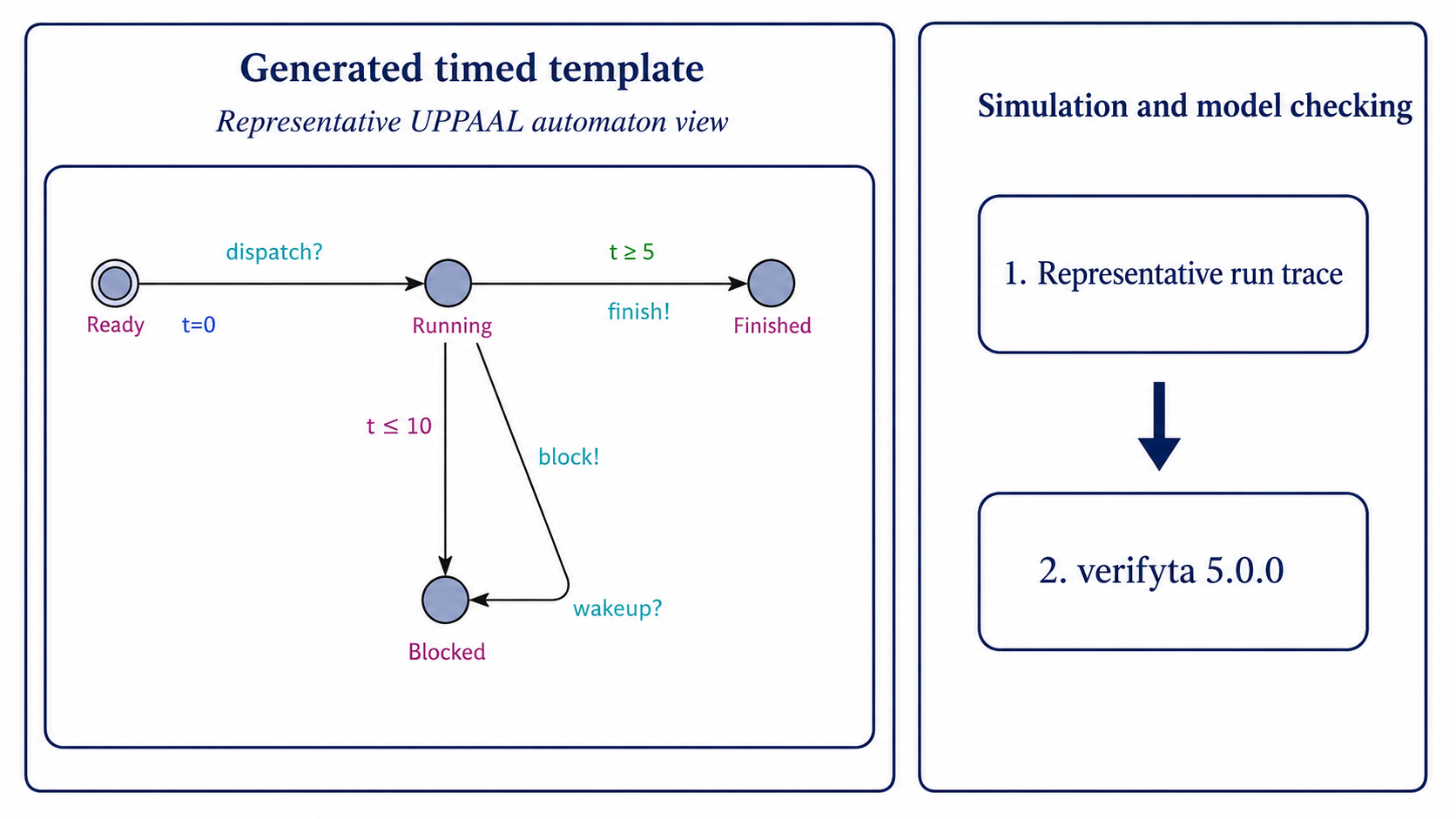}
  \caption{UPPAAL model and verification workflow.}
  \label{fig:uppaal-evidence}
\end{figure}

\subsection{Limitations and Validity}

Primary-model results use frozen completions re-used across replays; the full 91-family GLM-5.3 replication (Table~\ref{tab:multimodel}) covers the second model family, but two families of models do not establish model-independent calibration. The suites are development-grade: family labels come from a deterministic projection pending independent annotation review. External validity is limited to three evaluated protocols and 91 families; the cue gate generalizes at the pattern level and its recall is bounded by public cue coverage (Table~\ref{tab:cue-transfer}). Construct validity inherits the released annotation schema; checker soundness is relative to the finite RuleIR, and Section~\ref{sec:corruption} bounds the damage of its two error classes. The oracle arm reveals a benchmark artifact of its own: recorded minimal certificates omit catalog evidence, so the oracle is a lower-style ceiling on certificate-only retrieval rather than a true upper bound. FDR is the complement of precision and is retained only for continuity with prior reporting; unsafe-warrant behavior is measured on the authority suite. The UPPAAL case study validates compilation, not extraction. Broader protocols, independent annotation, and calibration remain future work.

\section{Conclusion}

LASER reframes protocol behavior recovery as defeasible certificate search rather than local clause extraction. Its central contribution is the interface between learned acquisition and symbolic warrant: typed routes gather sparse, mutually dependent evidence; policies decide where to search; and an occurrence-aware checker retains exclusive authority over support, conflict, and correction applicability. The revision strengthens this claim with three audit mechanisms: family-paired significance tests, a frozen-completion attribution replay showing the remote-reference gains require routing and typed expansion jointly, and an authority-corruption replay showing that source-link errors degrade to abstention while the single unsafe-warrant channel is localized to authority-status records. Unresolved errata and uncertain authority remain explicit rather than becoming unsupported transitions.

\begin{table}[!t]
\centering
\small
\setlength{\tabcolsep}{3.5pt}
\renewcommand{\arraystretch}{1.03}
\begin{tabular}{llrr}
\toprule
Evaluation & Metric & Comparator & LASER\\
\midrule
Applicable errata & Closed@1 & 0/25 & \textbf{25/25}\\
 & Mean expansions & 1.87 & \textbf{1.00}\\
Unresolved errata & Unsafe warrant & 0/6 & 0/6\\
Authority claims & Correct claims & 38/40 & \textbf{40/40}\\
 & Correct pairs & 18/20 & \textbf{20/20}\\
Held authority & Unsafe warrant & 0/8 & 0/8\\
\bottomrule
\end{tabular}
\caption{Outcomes beyond BehaviorCheck.}
\label{tab:cross-suite-results}
\end{table}

\begingroup
\small
\bibliography{references}
\endgroup
\end{document}